\documentclass[graybox]{svmult}

\usepackage{mathptmx}       
\usepackage{helvet}         
\usepackage{courier}        
\usepackage{type1cm}        
\usepackage{makeidx}         
\usepackage{graphicx}        
\usepackage{multicol}        
\usepackage[bottom]{footmisc}

\makeindex             

\global\arraycolsep=2pt
\usepackage{pstricks, pst-node, pst-text, pst-3d,epsfig}
\usepackage{amsmath}
\usepackage{epsfig}
\newcommand{\be}{\begin{equation}}
\newcommand{\ee}{\end{equation}}
\newcommand{\bea}{\begin{eqnarray}}
\newcommand{\eea}{\end{eqnarray}}
\newcommand{\ben}{\begin{equation*}}
\newcommand{\een}{\end{equation*}}
\newcommand{\bean}{\begin{eqnarray*}}
\newcommand{\eean}{\end{eqnarray*}}
\newcommand{\nn}{\nonumber}
\newcommand{\tr}{\mbox{tr}\,}
\newcommand{\Tr}{\mbox{Tr}\,}
\newcommand{\bGamma}{\mbox{\boldmath{$\Gamma$}}}

\begin{document}

\title*{Multiple Scattering: Dispersion, Temperature Dependence, and
Annular Pistons}
\titlerunning{Multiple Scattering}
\author{Kimball A. Milton, Jef Wagner, Prachi Parashar, In\'es Cavero-Pel\'aez, 
Iver Brevik, and Simen \AA. Ellingsen}

\institute{K. A. Milton, J. Wagner, P. Parashar \at H.L. Dodge Department of Physics and Astronomy,
University of Oklahoma, Norman, OK 73019 USA \email{milton@nhn.ou.edu, wagner@nhn.ou.edu, prachi@nhn.ou.edu} 
\and
I. Cavero-Pel\'aez \at Department of Theoretical Physics, Zaragoza University,
50009 Zaragoza, Spain \email{icaverop@gmail.com}
\and I. Brevik, S. \AA. Ellingsen \at
 Department of Energy and Process Engineering, Norwegian
University of Science and Technology, N-7491 Trondheim, Norway 
\email{iver.h.brevik@ntnu.no, simen.a.ellingsen@ntnu.no}}

\authorrunning{Milton et al.}

%
%
\maketitle

\abstract{We review various applications of the multiple scattering approach
to the calculation of Casimir forces between separate bodies, including 
dispersion, wedge geometries, annular pistons, and temperature
dependence.  Exact results are obtained in many cases.}

\section{Quantum Vacuum Energy}
\label{sec:1}
Quantum vacuum energies, or Casimir energies, are
important at all energy scales, from subnuclear to cosmological.
Applications are starting to appear in nanotechnology.
Furthermore it is most likely that the source of dark energy
that makes up some 70\% of the energy budget of the universe is quantum
vacuum fluctuations.  In particular,
the 7-year WMAP data is completely consistent with
the existence of a cosmological constant \cite{Komatsu:2010fb},
\be
w\equiv \frac{p}\rho=-1.10\pm0.14 (68\%\mbox{\,CL}),
\ee
which is precisely what would be expected if dark energy arose from this
source \cite{Weinberg:1988cp}.
Finally, zero-point fluctuations may be the most fundamental aspect of quantum 
field theory.

\section{Multiple-Scattering Formulation}
\label{sec:2}

The multiple scattering formulation 
is easiest stated for a scalar field, which is rather `easily' generalized to
electromagnetism.  For example, see Ref.~\cite{Kenneth:2007jk}.  
Vacuum energy is given by the famous trace-log formula,
\be
E=\frac{i}2\mbox{Tr}\,\ln G\to \frac{i}2\mbox{Tr}\,\ln G G_0^{-1},
\ee
where in terms of the background potential $V$,
\be
(-\partial^2+V)G=1, \quad -\partial^2 G_0=1. \ee

Now we define the $T$-matrix,
\be
T=S-1=V(1+G_0V)^{-1},\ee
and if the potential has two disjoint parts,
$V=V_1+V_2$,
it is easy to derive
the interaction between the two bodies (potentials):
\begin{subequations}
\bea
E_{12}&=& -\frac{i}{2}\mbox{Tr}\ln(1-G_0T_1G_0T_2)\\
&=&-\frac{i}{2}\mbox{Tr}\ln(1-V_1G_1V_2G_2),\label{ms2}\eea
\end{subequations}
where $G_i=(1+G_0V_i)^{-1}G_0$, $i=1,2$, and likewise $T_i$ refers
to $V_i$.

\section{Quantum Vacuum Energy---Dispersion}

Perhaps not surprisingly in retrospect, we find that the
usual dispersive form of the electromagnetic energy \cite{embook}
\be
U=\frac12\int (d\mathbf{r})
\int_{-\infty}^\infty \frac{d\omega}{2\pi}
\left[\frac{d(\omega\varepsilon)}{d\omega}E^2(\mathbf{r})
+H^2(\mathbf{r})\right]\label{dispen}
\ee
must be used,
which, quantum mechanically, corresponds to the vacuum energy form
\be
\mathcal{E}=-\frac{i}{2}\int(d\mathbf{r})\int\frac{d\omega}{2\pi}
\left[2\varepsilon\tr\bGamma+
\omega\frac{d\varepsilon}{d\omega}\tr\bGamma\right],\label{dispenergy}
\ee
in terms of the Green's dyadic $\bGamma$.
This result follows directly from the trace-log formula for the vacuum
energy
\be
\mathcal{E}=\frac{i}2\int\frac{d\omega}{2\pi}\Tr\ln\bGamma,\ee
and is equivalent to the variational statement \cite{Schwinger:1977pa}
\be
\delta\mathcal{E}=\frac{i}2\int\frac{d\omega}{2\pi}\Tr\delta
\varepsilon\bGamma.
\ee

From the energy, precisely because the dispersive derivative terms are
present, we recover the Lifshitz formula for the energy per area between 
parallel dielectric slabs, with permittivity $\varepsilon_{1,2}$,
separated by a medium of permittivity $\varepsilon_3$ of thickness $a$,
\be
\frac{\mathcal{E}}A=\frac1{4\pi^2}\int_0^\infty d\zeta\int_0^\infty dk\, k\left[
\ln\left(1-r_{\rm TE}r_{\rm TE}'e^{-2\kappa_3 a}\right)+
\ln\left(1-r_{\rm TM}r_{\rm TM}'e^{-2\kappa_3 a}\right)\right],\label{lifshitz}
\ee
with $\kappa_i=\sqrt{k_\perp^2+\zeta^2\varepsilon_i}$, $\zeta=-i\omega$ being
the imaginary frequency.
The TE reflection coefficients are given by
\be
r_{\rm TE}=\frac{\kappa_3-\kappa_1}{\kappa_3+\kappa_1},\quad
r'_{\rm TE}=\frac{\kappa_3-\kappa_2}{\kappa_3+\kappa_2},
\ee
while the TM coefficients are obtained from these by the substitution $\kappa_a\to
\bar\kappa_a=\kappa_a/\epsilon_a$.
For further details of this calculation, see Ref.~\cite{Milton:2010yw}.

\section{Noncontact Gears}
\begin{figure}
\sidecaption
\includegraphics[width=70mm,scale=1.0]
{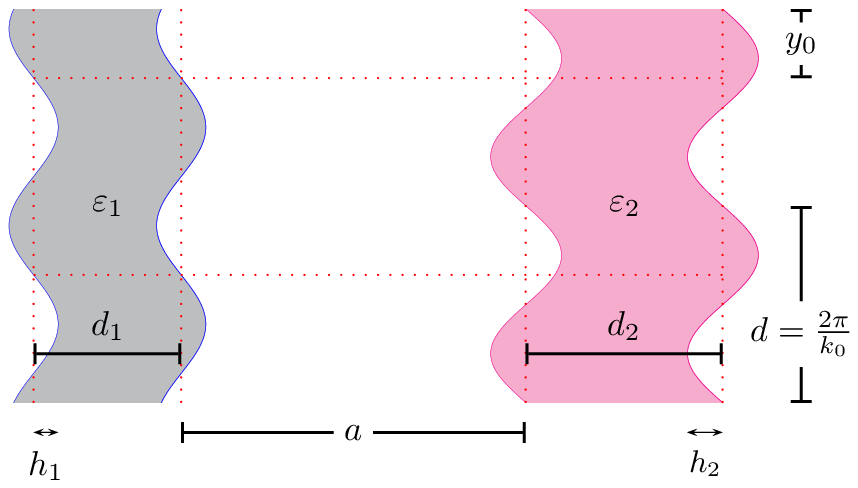}
\caption{Parallel dielectric slabs with sinusoidal corrugations.}
\label{corru}
\end{figure}

The program of calculating the quantum vacuum lateral
force between corrugated surfaces and gears has been
under active development.  The electromagnetic situation
of currugated dielectric slabs is illustrated in Fig.~\ref{corru}.
For details see Ref.~\cite{Parashar:2010yj}.

\begin{figure}
\sidecaption
\includegraphics[width=2.5in,scale=1.0]
{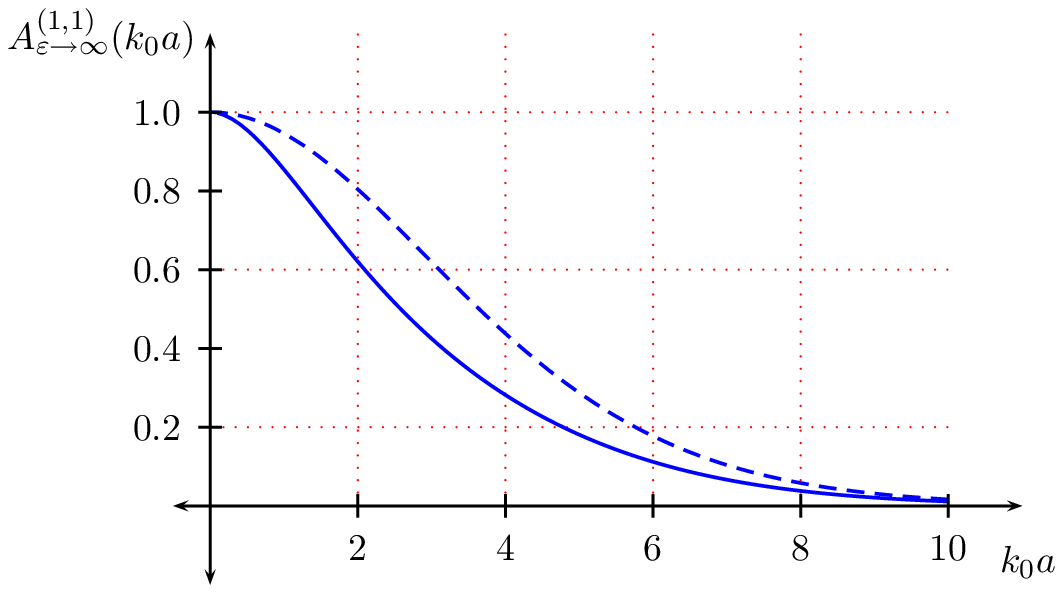}
\caption{Plot of $A^{(1,1)}_{\varepsilon\rightarrow\infty}(k_0a)$
versus $k_0a$. The dotted curve represents 2 times the Dirichlet case.}
\label{Aem11-cond-versus-t0}
\end{figure}

In the conductor limit ($\varepsilon_i\rightarrow\infty$) 
and for the case of sinusoidal corrugations described by
$h_1(y)=h_1\sin [k_0(y+y_0)]$ and
$h_2(y)=h_2\sin [k_0y]$ the lateral force can be evaluated to be
in first order in $h_1/a$ and $h_2/a$
\begin{equation}
F^{(2)}_{\varepsilon\rightarrow\infty}
= 2k_0a\, \sin (k_0y_0) \left|F^{(0)}_\text{Cas}\right|
\frac{h_1}{a} \frac{h_2}{a}
A^{(1,1)}_{\varepsilon\rightarrow\infty} (k_0a),
\end{equation}
where
\begin{equation}
A^{(1,1)}_{\varepsilon\rightarrow\infty} (t_0)
= \frac{15}{\pi^4}\int_{-\infty}^\infty dt \int_0^\infty \bar{s}d\bar{s}
\frac{s}{\sinh s} \frac{s_+}{\sinh s_+}
\left[ \frac{1}{2} + 
\frac{(s^2 + s_+^2 - t_0^2)^2}{8\,s^2s_+^2} \right],
\label{A-cond-2int}
\end{equation}
where $s^2=\bar{s}^2+t^2$ and $s_+^2=\bar{s}^2+(t+t_0)^2$.
The first term in  Eq.\,\eqref{A-cond-2int} corresponds to the 
Dirichlet scalar case~\cite{CaveroPelaez:2008tj}, which here corresponds 
to the E mode (referred to in Ref.~\cite{Emig:2003} as the TM mode).
We note that $A^{(1,1)}_{\varepsilon\rightarrow\infty} (0)=1$.
See Fig.~\ref{Aem11-cond-versus-t0} for the plot of
$A^{(1,1)}_{\varepsilon\rightarrow\infty}(k_0a)$ versus $k_0a$.
We observe that only in the proximity force approximation limit
$k_0a=0$  is the electromagnetic 
contribution twice that of the Dirichlet case, and in general 
the electromagnetic case is less than twice that of the Dirichlet case.
This result can be shown to coincide with the expression
found in Emig {\it et al.} \cite{Emig:2003} 
apart from an overall factor of 2,
which presumably is a transcription error.
The double integral representation in Eq.~\eqref{A-cond-2int}
is more useful for numerical evaluation than the single-integral
form given in Ref.~\cite{Emig:2003} because of the oscillatory 
nature of the function $\sin x/x$ in the latter. 
Generalization of these results are forthcoming.

\section{Wedge as generalization of cylinder}
\begin{figure}
\sidecaption
  \psfig{file=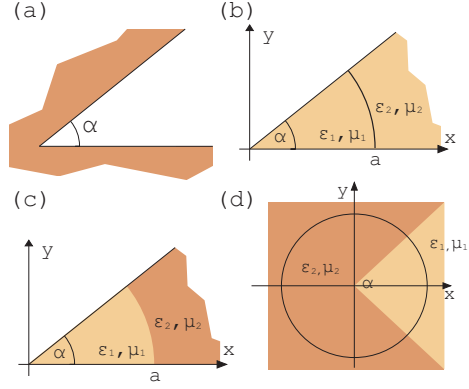, width = 2.5in}
  \caption{Wedge geometries. (a) The perfectly conducting wedge geometry.
(b) The geometry of a wedge intercut by a perfectly conducting cylindrical arc.
 (c) Wedge with magnetodielectric arc. (d) Diaphanous wedge in a perfectly
conducting cylindrical shell.}
  \label{fig:wedge}
\end{figure}

In a series of papers, we have considered variations on the wedge
geometry, such as a wedge defined by perfectly reflecting walls, intersected
with a concentric circular cylinder, the arc being either a perfect reflector
itself, or the boundary between two dielectric-diamagnetic regions.  Most
interesting is the case when the wedge itself is constructed as the interface
between two such media.  See Fig.~\ref{fig:wedge}.  In order to have a
tractable situation, we have considered the diaphanous or isorefractive condition
\be
\varepsilon_1\mu_1=\varepsilon_2\mu_2,
\ee
that is, the speed of light is the same in the two media.  (If that is not
done for the wedge, the differential equations are no longer separable.)
See Refs~\cite{Brevik:2009vf,Ellingsen:2009ff,Ellingsen:2010yj} for more detail.

\begin{figure}[tb]
\sidecaption[t]
  \includegraphics[width=2in]{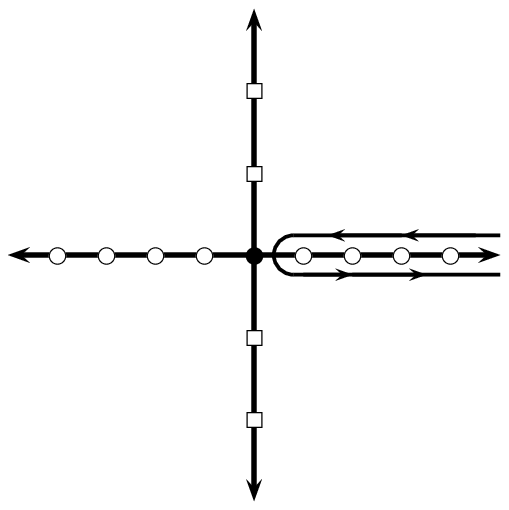}
  \caption{Contour of integration $\gamma$ for the argument principle. 
Shown also are singularities of the integrand
 along the real and imaginary $\nu$ axes.}
  \label{fig-gamma}
\end{figure}

Consider now case (d).
Using multiple scattering, or the  Kontorovich-Lebedev
transformation, we obtain the following implicit formula
for the eigenvalues for the order $\nu$ of the contributing cylindrical 
partial waves,
$D(\nu,\omega)=0,$ where
($r= $ reflection coefficient on wedge)
\bea
  D(\nu,\omega) &=& (1-e^{2\pi i\nu})^2- r^2(e^{i\nu(2\pi-\alpha)}-
e^{i\nu\alpha})^2 \nn\\
  &=& -4e^{2\pi i\nu}[\sin^2(\nu\pi)- r^2\sin^2(\nu(\pi-\alpha))],
\eea
which are selected by the ``argument principle,'' which is just the
Cauchy theorem applied to the contour $\gamma$ shown in Fig.~\ref{fig-gamma}.

\begin{figure}[tb]
\sidecaption
  \includegraphics[width=2.5in]{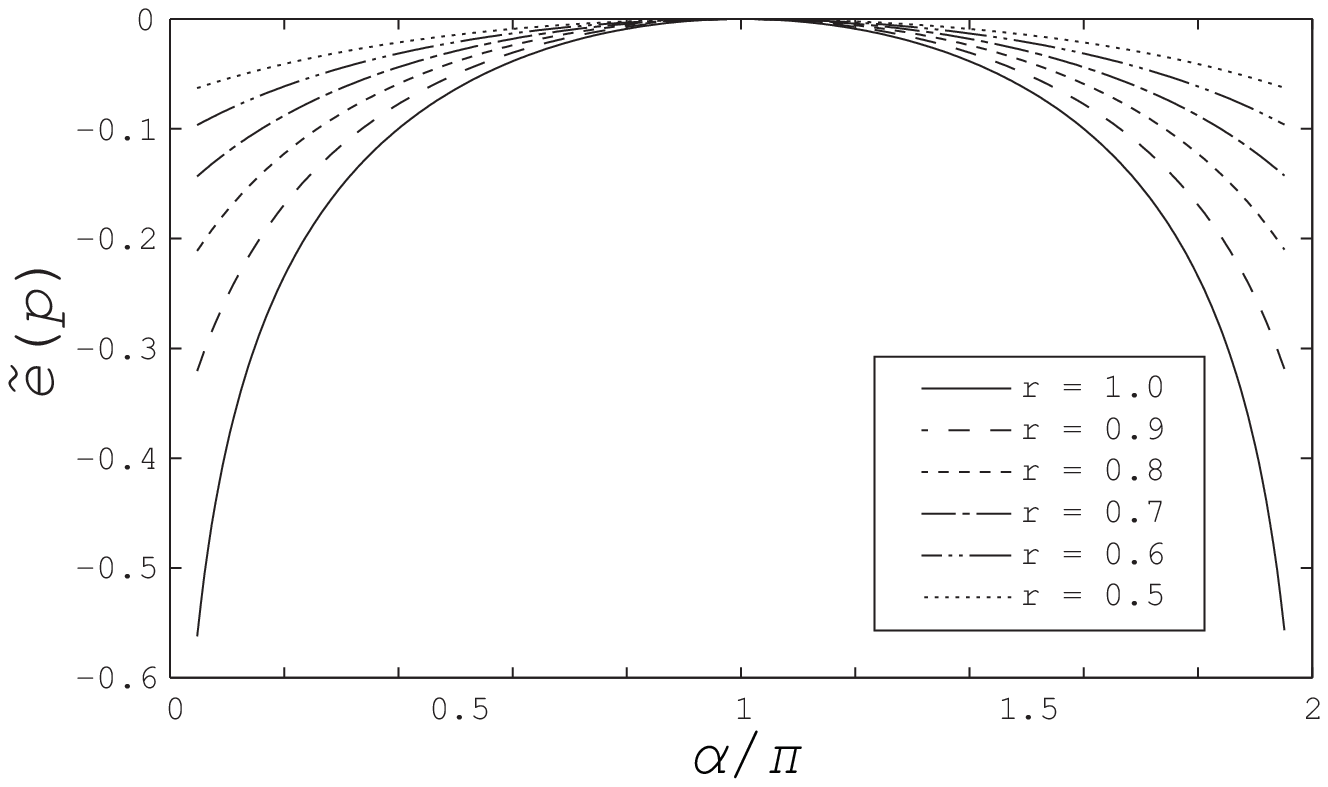}
  \caption{The function $\tilde{e}(p)$ 
plotted as a function of opening angle $\alpha$. }
  \label{fig_ep}
\end{figure}

In this way, we find the energy per length given by
$\tilde{\mathcal{E}} = \frac1{8\pi na^2}\tilde{e}(p)$, $p=\pi/\alpha$
 as shown in Fig.~\ref{fig_ep}.
Note that only for perfect reflectors does the energy diverge as the opening
angle approaches zero.

\section{Annular Piston---Semitransparent Plates}

\begin{figure}[tb]
\sidecaption
    \includegraphics[width=1.4in]{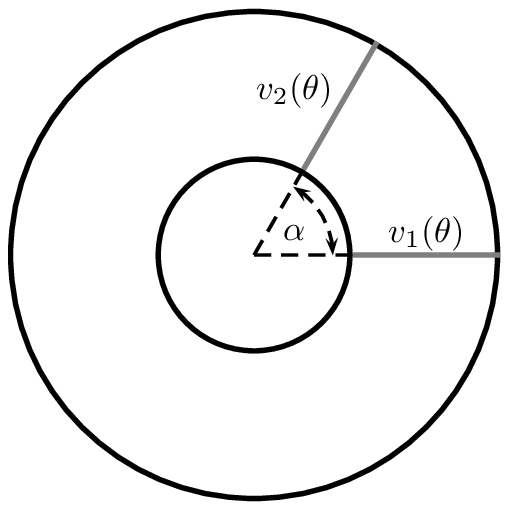}
    \caption{Two semitransparent plates in an annulus.}
\label{ann-piston}
\end{figure}
The wedge geometry may be generalized by considering
two semitransparent plates in a Dirichlet annulus, as shown in 
Fig.~\ref{ann-piston}.
We use multiple scattering in the angular coordinates,
and an eigenvalue condition in the radial coordinates; this
problem is equally well solvable
 with radial Green's functions, but this approach may be more
 generalizable.  This section is based on Ref.~\cite{Milton:2009bz}.

The Green's function $\mathcal{G}(\mathbf{r,r'})$ will satisfy the equation
\be
  \left[-\nabla^2-\omega^2+V(\mathbf{r})\right]\mathcal{G}(\mathbf{r,r'})=
  \delta(\mathbf{r-r'}),
\ee
while $\mathcal{G}^{(0)}$ has $V(\mathbf{r})=0$.
For the cylindrical
geometry of an annulus, the boundary conditions are $\mathcal{G}=0$ at $\rho=a$ and $\rho=b$,
where $a$ and $b$ are the inner and outer radii, respectively.  We take the
potential to be $V(\mathbf{r}) = v(\theta)/\rho^2$.
The corresponding Green's function is
\be
  \mathcal{G}(\mathbf{r,r'};\omega)=\int_{-\infty}^\infty
  \frac{d k}{2\pi}e^{i k(z-z')}
  \sum_\eta R_\eta(\rho;\omega,k)R_\eta(\rho';\omega,k)
 g_\eta(\theta,\theta'),
\ee
in terms of the separation constant $\eta$.
The normalized radial eigenfunctions appearing here are
\be\label{rad_de}
  \left[-\rho \frac{d}{d\rho} \rho \frac{d}{d\rho} -(\omega^2 -k^2)\rho^2
\right]  R_\eta(\rho;\omega,k)=\eta^2R_\eta(\rho;\omega,k),
\ee
with the boundary conditions $R_\eta(a;\omega,k)=R_\eta(b;\omega,k)=0$.
The reduced Green's function satisfies
\be
  \left[-\frac{d^2}{d\theta^2} +\eta^2+v(\theta)\right]g_\eta(\theta,\theta')=
  \delta(\theta-\theta'),
\ee
with periodic boundary conditions.

To obtain the radial functions,
we need the solution of the modified Bessel differential equation, of 
 imaginary order, which is
zero for $\rho=a$ for all values of $\eta$ and $\kappa$. An obvious solution is
\be
  \tilde{R}_\eta(\rho;\kappa)=K_{i\eta}(\kappa a)\tilde I_{i\eta}(\kappa \rho) 
- \tilde I_{i\eta}(\kappa a) K_{i\eta}(\kappa \rho)=
\tilde R_{-\eta}(\rho,\kappa),
\label{tilder}
\ee
where
\be
\tilde I_\nu=\frac12(I_\nu+I_{-\nu}).
\ee
The eigenvalues are given by the zeros of
$D(\eta)=\tilde R_\eta(b;\kappa)$. We don't need the explicit eigenfunctions here.
\subsection{Reduced Green's Function}

The free angular reduced Green's function is given by
\be
  g^{(0)}_\eta(\theta,\theta')=\frac{1}{2\eta}\left(-\sinh \eta|\theta-\theta'|
  +\frac{\cosh \eta \pi}{\sinh \eta \pi} \cosh \eta |\theta-\theta'|\right).
\ee
For a single potential $v(\theta)=\lambda\delta(\theta-\alpha)$ 
for $\theta, \theta'\in[\alpha,2\pi+\alpha]$, the
reduced Green's function is
\bea
g_\eta(\theta,\theta')&=&
 \frac{1}{2\eta}\bigg(-\sinh \eta |\theta-\theta'|
  +\frac{2\eta \cosh \eta \pi \cosh \eta |\theta-\theta'|}
{2\eta \sinh \eta \pi +\lambda \cosh \eta \pi}\nn\\
&& \mbox{}- \lambda\frac{
    \cosh \eta ( 2\pi +2 \alpha -\theta -\theta')
    -\cosh 2 \eta \pi \cosh \eta |\theta-\theta'|}
   {[2\eta \sinh \eta \pi +\lambda \cosh \eta \pi]2\sinh \eta \pi}\bigg).
\label{1pgf}
\eea

\subsection{Two Semitransparent Planes}
Now we look at the interaction energy between two semitransparent planes,
as illustrated in Fig.~\ref{ann-piston}.
Since it is nontrivial to work out the Green's function for two
potentials, it is easiest to use the
multiple-scattering formalism (\ref{ms2})
\be
  E=\frac{1}{2i}\int_{-\infty}^\infty \frac{d \omega}{2\pi}
  \Tr \ln ( 1- \mathcal{G}^{(1)}V_1\mathcal{G}^{(2)}V_2 ).
\ee
The subscripts on the $V$s represent the potentials $V_1(\mathbf{r})=
\lambda_1\delta( \theta ) / \rho^2$, and $V_2(\mathbf{r}) =
\lambda_2 \delta( \theta- \alpha ) / \rho^2$.
The Green's functions with superscript $(i)$
represent the interaction with only a single
potential $V_i$. 
From this we obtain a simplified form of the interaction energy:
\be
  \mathcal{E}=\frac{1}{4\pi}\int_0^\infty \kappa \,d \kappa
  \sum_\eta \ln \left( 1- \tr g_\eta^{(1)}v_1 g_\eta^{(2)}v_2 \right),
\ee
where $g_\eta^{(i)}$ are given in Eq.~(\ref{1pgf}). Then
\be
  \tr g_\eta^{(1)}v_1 g_\eta^{(2)}v_2  =\frac{\lambda_1 \lambda_2
    \cosh^2 \eta (\pi-\alpha)}
  {\left(2\eta \sinh \eta \pi +\lambda_1 \cosh \eta \pi\right)
    \left(2 \eta \sinh \eta \pi +\lambda_2 \cosh \eta \pi \right) }.
\ee

Using the argument principle to determine the angular eigenvalues,
we get the following expression for the energy for an annular Casimir
piston,
\bea
\mathcal{E}
&=&\frac1{8\pi^2 i}\int_0^\infty \kappa d \kappa
  \int\limits_\gamma \!d \eta\!
  \frac\partial{\partial\eta} \ln \left[
    K_{i\eta}(\kappa a)\tilde I_{i\eta}(\kappa b) -
    \tilde I_{i\eta}(\kappa a) K_{i\eta}(\kappa b)\right]\nn\\
 && \times\ln \left( 1- \frac{\lambda_1 \lambda_2 \cosh^2 \eta (\pi-\alpha)
/\cosh^2\eta\pi}
  {\left(2\eta \tanh \eta \pi +\lambda_1\right)
    \left(2 \eta \tanh \eta \pi +\lambda_2  \right) } \right).
\eea
The contour of integration for the argument principle is again given in
Fig.~\ref{fig-gamma}.

\begin{figure}
\sidecaption
\psfig{file=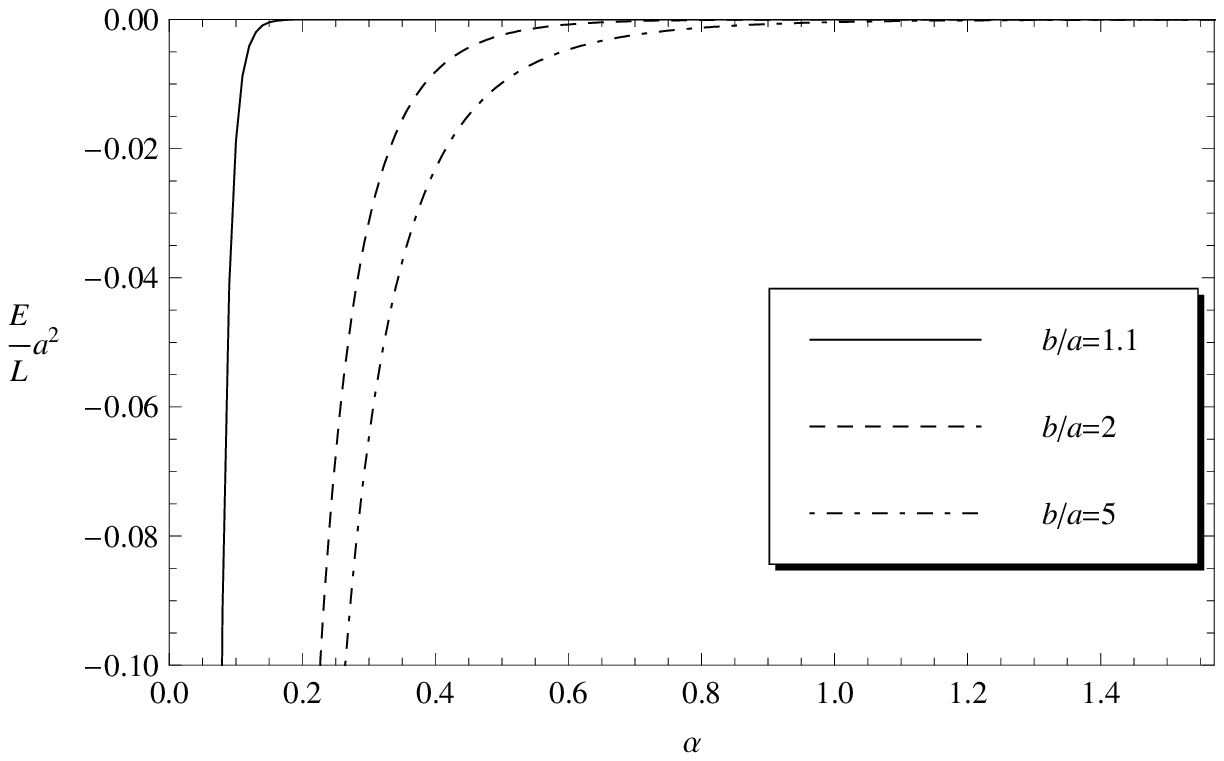,width=3in}
\psfig{file=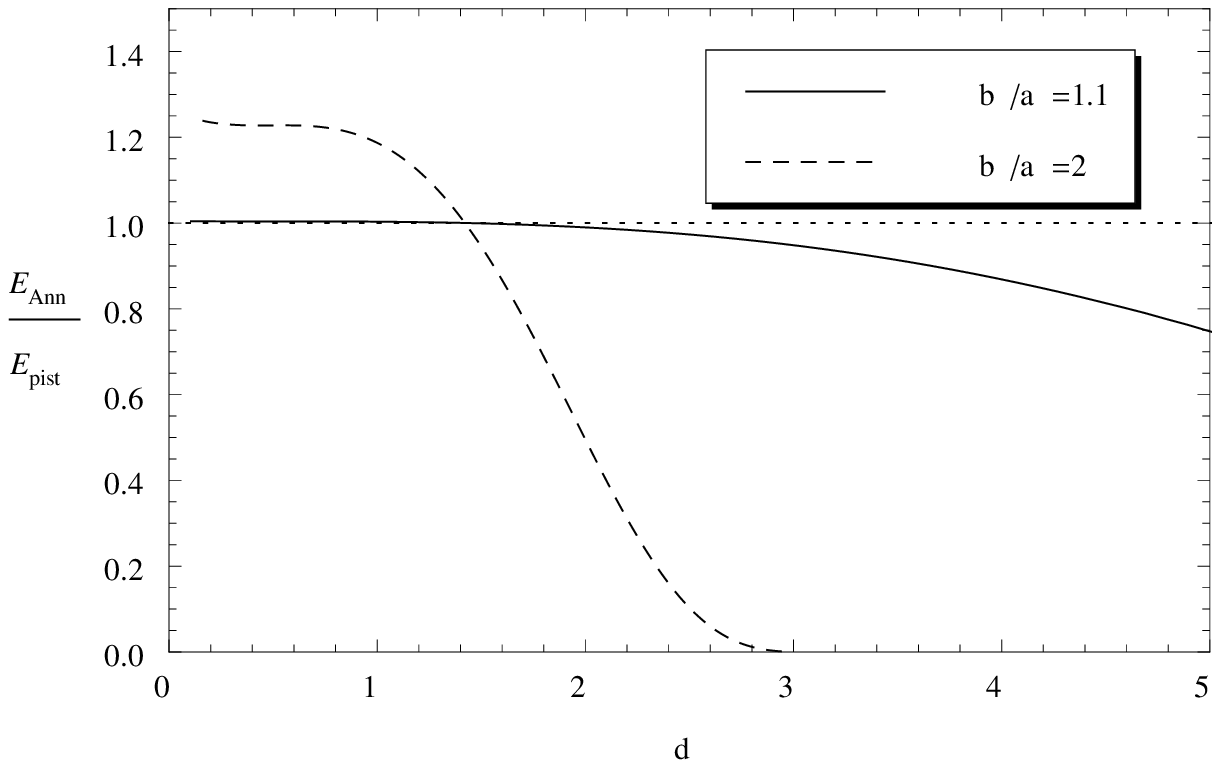,width=3in}
\caption{\label{figap} Energy/length for an annular piston as function of 
angle (top),
and compared to the energy/length for a rectangular piston (bottom).}
\end{figure}

This formula can actually be used to evaluate the energy of interaction
between the two planes of the piston, by distorting the $\eta$ contour to lines
making angles of $\pm\pi/4$ with respect to the real axis.  The results
are shown in Fig.~\ref{figap}.
In Fig.~\ref{figap} we define $ d=\frac{b+a}2 \sin\frac\alpha2,$
and the plateaus seen in the second figure
may be understood from the proximity force
approximation,
\be
\frac{\mathcal{E}_{\rm PFA}}{\mathcal{E}_\|}=\frac1{16}\frac{b^2}{a^2}
\left(1+\frac{a}b\right)^4,\ee
in comparison to the interaction between infinite parallel plates.

\section{Applications of Multiple Scattering}
\begin{figure}
\sidecaption
\epsfig{file=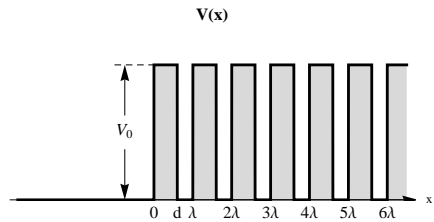,height=3cm}
\caption{ A semi-infinite array of periodic potentials.
The exact CP force between an atom and this array may be calculated.}
\label{fig-pp}
\end{figure}

As an illustration of practical calculations using the multiple
scattering machinery, we illustrate in Fig.~\ref{fig-pp}
a semi-infinite array of
periodic potentials, such as a array of dielectric slabs, for
which the exact Casimir-Polder force with an atom to the left may be
calculated \cite{cas2009}.

\subsection{Casimir-Polder Force}
Consider an atom, of polarizability $\alpha(\omega)$, a distance $Z$ to the
left of the array.  The Casimir-Polder energy is
\be
E=-\int_{-\infty}^\infty
 d\zeta\int\frac{d^2k}{(2\pi)^2}\alpha(i\zeta)\mbox{tr}\,{\bf g}(Z,Z),
\ee
where apart from an irrelevant constant the trace of the Green's function is
\be \mbox{tr}\,{\bf g}(Z,Z)\to
\frac1{2\kappa}\left[-\zeta^2\mathcal{R}^{\rm TE}+(\zeta^2+2k^2)\mathcal{R}^{
\rm TM}\right]e^{-2\kappa|Z|}.
\ee
Here the reflection coefficients are those for the entire array ($a$ is the
distance between the potential slabs),
\be
\mathcal{R}=\frac1{2R}\bigg[e^{2\kappa a}+R^2-T^2-\sqrt{\left(e^{2\kappa a}
-R^2-T^2\right)^2-4 R^2T^2}\bigg].
\ee
If the potentials consist of dielectric slabs, with dielectric constant
$\varepsilon$ and thickness $d$, the TE reflection and transmission
coefficients for a single slab are ($\kappa'=\sqrt{\varepsilon\zeta^2+k^2}$)
\begin{subequations}
\bea
R^{\rm TE}&=&\frac{e^{2\kappa' d}-1}{\left(\frac{1+\kappa'/\kappa}{1-\kappa'
/\kappa}\right)e^{2\kappa'd}-\left(\frac{1-\kappa'/\kappa}{1+\kappa'/\kappa}
\right)},\\
T^{\rm TE}&=&\frac{4(\kappa'/\kappa) e^{\kappa' d}}
{(1+\kappa'/\kappa)^2e^{2\kappa'd}-(1-\kappa'/\kappa)^2}.
\eea
\end{subequations}
The TM reflection and transmission coefficients are obtained by replacing,
 except in the exponents,
$\kappa'\to\kappa'/\varepsilon$.  (Multilayer potentials have been
discussed extensively in the past, see, for example, Refs.~\cite{zhou,tomas,
casbook09,parsegian}.)

\begin{figure}
\sidecaption[t]
\epsfig{file=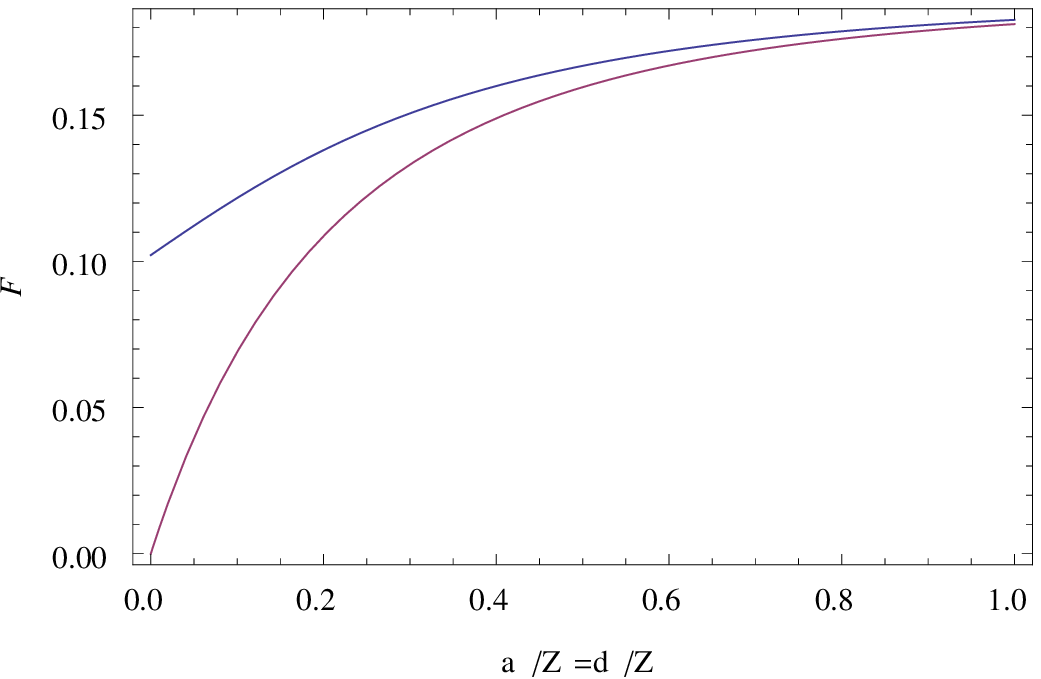,height=4cm}
\caption{\label{cpfig} Casimir-Polder energy between a semi-infinite array
of dielectric slabs with $\varepsilon=2$, compared to the energy (lower
curve) if only one slab were present.  Here we have assumed that the
spacing between the slabs and the widths of the slabs are equal.}
\end{figure}

\begin{figure}
\sidecaption
\epsfig{file=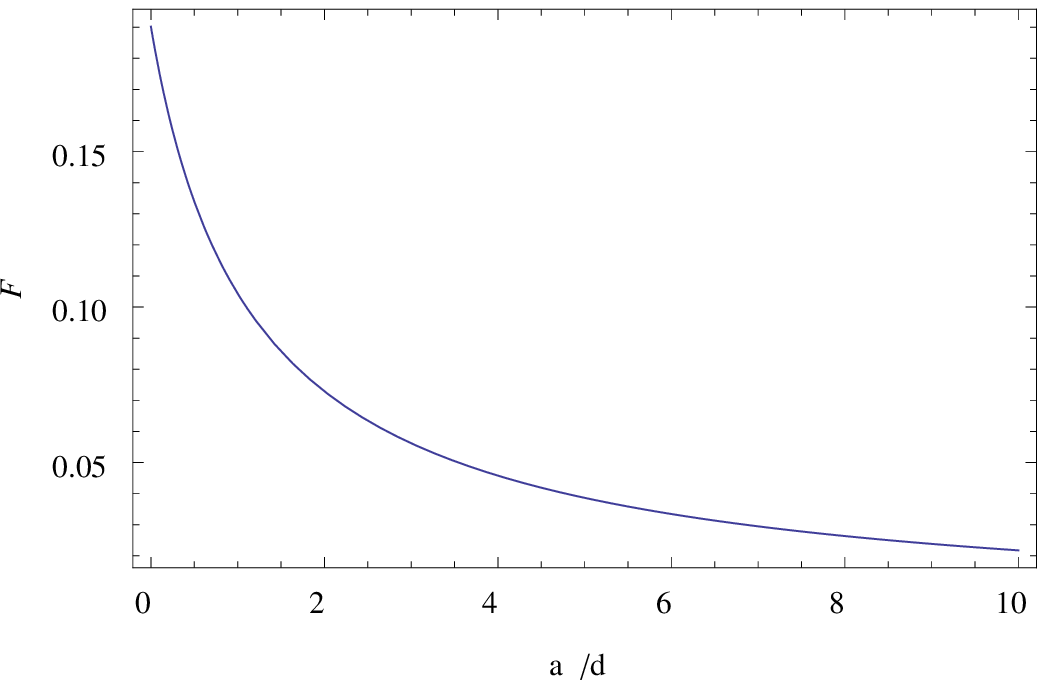,height=4cm}
\caption{\label{ztoinfty} Casimir-Polder energy for large distances from
the array, as a function of the ratio $a/d$, where $a$ is the
distance between the dielectric slabs in the array, and $d$ is
the thickness of each slab.  Here $\varepsilon=2$.}
\end{figure}

For example, in the static limit, where we disregard the frequency
dependence of the polarizability,
\be
E=-\frac{\alpha(0)}{2\pi}\frac1{Z^4}F(a/Z,d/Z).
\ee
This is compared with the single slab result in Fig.~\ref{cpfig}.
It is interesting to consider the 
$Z\to\infty$ limit, which is shown in Fig.~\ref{ztoinfty}.
When $a/d\to0$ we recover the bulk limit.
Such results apparently will have applications to experiment rather
soon \cite{keil}.

\section{Exact Temperature Results}

The scalar Casimir energy between two weak nonoverlapping potentials
$V_1({\bf r})$ and $V_2({\bf r})$ at temperature $T$ is \cite{ext}
\be
E_T=-\frac{T}{32\pi^2}\int(d\mathbf{r})(d\mathbf{r'})V_1(\mathbf{r})
V_2(\mathbf{r'})\frac{\coth 2\pi T|\mathbf{r-r'}|}{|\mathbf{r-r'}|^2}.
\label{ET}
\ee

\subsection{Exact Proximity Force Approximation}
From Eq.~(\ref{ET}) we find that
the energy between a semitransparent plane and an arbitrarily curved
nonintersecting semitransparent surface is for weak coupling 
\be
E_T=-\frac{\lambda_1\lambda_2 T}{16\pi}\int dS\int_{2\pi T z(S)} dx
\frac{\coth x}x,\label{etcoth}
\ee
where the area integral is over the curved surface.  Here $z(S)$ is
the distance between the plates at a given point on the surface $S$.
Equation (\ref{etcoth})  is precisely what one means by the proximity force approximation:
\be
E_{\rm PFA}=\int dS\mathcal{E}_\|(z(S)),
\ee
as noted by Decca et al.~\cite{Decca:2009fg}.  See also Ref.~\cite{Dalvit:2009gw}.

\subsection{Interaction Between Semitransparent Spheres}

We can, for weak scalar coupling, compute the energy between
two spheres of radius $a$ and $b$, whose centers are separated
by a distance $R$:
\bea
E_T&=&-\frac{\lambda_1\lambda_2}{16 \pi}\frac{ab}R
\bigg\{\ln\frac{1-(a-b)^2/R^2}{1-(a+b)^2/R^2}+f(2\pi T(R+a+b))
+f(2\pi T(R-a-b))\nn\\
&&\quad\mbox{}-f(2\pi T(R-a+b))
-f(2\pi T(R+a-b))\bigg\},
\eea
where $f(y)$ for $y<\pi$ is given by the power series,
\be
f(y)=\sum_{n=1}^\infty \frac{2^{2n}B_{2n}}{2n(2n-1)(2n)!}y^{2n},
\label{ltexp}
\ee
which is
obtained from the differential equation
\be
y\frac{d^2}{dy^2}f(y)=\coth y-\frac1y,\quad f(0)=f'(0)=0.\label{diffeq}
\ee

\begin{figure}
\sidecaption
\includegraphics[width=2.5in]{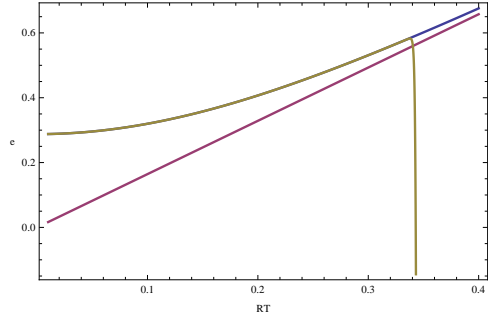}
\caption{Comparison between the general and high temperature forms
of the energy, as a function of $RT$. Energies are shown for $a=b=R/4$.
The high temperature result is linear in $T$. Also shown is the power
series expansion  truncated at 200 terms.
which diverges in this case at $RT=1/3$.  Plotted is
$e=-16\pi R E/(\lambda_1\lambda_2 a^2)$.}
 \label{figx}
 \end{figure}

Results for the energy obtained by solving this differential equation 
are shown in Fig.~\ref{figx}.  For further
details see Ref.~\cite{ext}.

\subsection{Mean distances between spheres}
Encountered in the above calculation are
mean powers of distances between spheres as defined by
\be
\int d\Omega\,d\Omega' |\mathbf{r-r'}|^p=(4\pi)^2 R^p P_p({\hat a},{\hat b}),
\ee for
spheres, of radii $a$ and $b$, respectively, separated by a center-to-center
distance $R$.  Here $\hat a=a/R$ and
$\hat b=b/R$, and $P_p({\hat a},{\hat b})$ can in general be represented
by the infinite series
\be
P_p(\hat a,\hat b)=\sum_{n=0}^\infty \frac{2}{(2n+2)!}\frac{\Gamma(2n-p-1)}
{\Gamma(-p-1)}Q_n(\hat a,\hat b).
\ee
Here the homogeneous polynomials $Q_n$ are
\begin{subequations}
\bea
Q_0&=&1,\\
Q_1&=&2({\hat a}^2+{\hat b}^2),\\
Q_2&=&3{\hat a}^4+10{\hat a}^2{\hat b}^2+3{\hat b}^4,\\
Q_3&=&4{\hat a}^6+28{\hat a}^4 b^2+28{\hat a}^2{\hat b}^4+4{\hat b}^6.
\eea
\end{subequations}
Here in general,
\be
Q_n=\frac12\sum_{m=0}^n{2n+2\choose 2m+1}{\hat a}^{2(n-m)}{\hat b}^{2m}.
\ee  There is also a recursion relation,
\be
P_{p-1}({\hat a},{\hat b})=\frac{R^{-p}}{1+p}\frac\partial{\partial R}
R^{1+p}P_p(\hat a,\hat b),\label{rr}
\ee
since $Q_n$ is homogeneous in $R$ of degree $-2n$.

For integer $p>-2$,
$P_p$ is a polynomial of degree $2\lceil p/2\rceil$,
and we can immediately find
\bea
P_p(\hat a,\hat b)&=&\frac1{4{\hat a}{\hat b}}\frac1{(p+2)(p+3)}
\left[(1+\hat a+\hat b)^{p+3}\right.\nn\\
&&\mbox{}+(1-\hat a-\hat b)^{p+3}
-\left.(1-\hat a+\hat b)^{p+3}-(1+\hat a-\hat b)^{p+3}\right],
\label{closedf}
\eea
Although this was derived for integer $p$ it actually holds for
all values of $p$.

For example, when $p$ is a negative integer, we have the explicit
forms, which are obtained from Eq.~(\ref{closedf}) by taking the appropriate
limit:
\begin{subequations}
\bea
P_{-1}&=&1,\quad{\mbox{Newton's theorem}},\\
P_{-2}&=&\frac1{4{\hat a}{\hat b}}\left[\ln\frac{1-({\hat a}+{\hat b})^2}
{1-({\hat a}-{\hat b})^2}+\hat a\ln\frac{(1+\hat b)^2-{\hat a}^2}
{(1-\hat b)^2-{\hat a}^2}
+\hat b\ln\frac{(1+\hat a)^2-{\hat b}^2}
{(1-\hat a)^2-{\hat b}^2}\right]
,\\
P_{-3}&=&-\frac1{4{\hat a}{\hat b}}\ln\frac{1-({\hat a}+{\hat b})^2}
{1-({\hat a}-{\hat b})^2},\\
P_{-4}&=&\frac1{[1-({\hat a}+{\hat b})^2][1-({\hat a}-{\hat b})^2]}.
\eea
\end{subequations}
and further expressions, which can be obtained by use of Eq.~(\ref{rr}),
may be readily verified.

\section{Conclusions}
 The multiple scattering formalism can be used to find numerical
results effectively in many situations, as we have seen in this
outline. Weak coupling results are exact and often given in closed form.
The method can also be used to extract not only interaction energies but
self energies, as described in Ref.~\cite{Milton:2010qr}.


%
\begin{acknowledgement}
We thank the US National Science Foundation and the US Department of Energy
for partial support of this research.  We thank Elom Abalo, Nima Pourtolami,
and K. V. Shajesh
for collaborative assistance.  We dedicate this paper to Emilio Elizalde.
\end{acknowledgement}
%

%
%
%

\end{document}